\documentclass[useAMS,usenatbib]{mn2e}
\usepackage{graphicx}
\usepackage{epsfig}
\usepackage{float}
\usepackage{amsmath}
\usepackage{natbib}
\title[X-ray bursts as a probe of the corona: 4U 1636-536]{X-ray bursts as a probe of the corona: \\ the case of XRB 4U 1636-536}

\author[L. Ji, et al.]
{\parbox{\textwidth}{Long Ji,$^{1}$\thanks{E-mail:  jilong@ihep.ac.cn (LJ); szhang@ihep.ac.cn (SZ); chenyp@ihep.ac.cn (YPC)}
Shu Zhang,$^{1}$\footnotemark[1]
YuPeng Chen,$^{1}$\footnotemark[1]
Shuang-Nan Zhang,$^{1}$
Diego F. Torres,$^{2,3}$
Peter Kretschmar,$^{4}$
Masha Chernyakova,$^{5}$
Jian Li$^{1}$, and
Jian-Min Wang$^{1,6}$}\vspace{0.4cm}\\
\parbox{\textwidth}{$^{1}$Laboratory for Particle Astrophysics, Institute of High Energy Physics, Beijing 100049, China\\
$^{2}$Instituci\'o Catalana de Recerca i Estudis Avan\c cats (ICREA), 08010 Barcelona, Spain\\
$^{3}$Institute of Space Sciences (IEEC-CSIC), Campus UAB, Torre C5, 2a planta, 08193 Barcelona, Spain\\
$^{4}$ISOC, ESA/ESAC, Urb. Villafranca del Castillo, PO Box 50727, 28080 Madrid, Spain\\
$^{5}$School of Physical Sciences, Dublin City University, Glasnevin, Dublin 9, Ireland\\
$^{6}$Theoretical Physics Center for Science Facilities (TPCSF), CAS, 19(B) Yuquan Road, Beijing 100049, China\\
}}

\begin{document}
\date{Accepted by MNRAS in April 2013}
\pubyear{2012}
\maketitle
\label{firstpage}
\begin{abstract}
To investigate the possible cooling of the corona by soft X-rays bursts, we have studied 114 bursts embedded in the known X-ray evolution of 4U 1636-536.
We have grouped these bursts according to the { ratio of the flux in the 1.5--12 keV band with respect to that in the 15--50 keV band, as monitored by RXTE/ASM and Swift/BAT, respectively. We have detected a shortage at hard X-rays while bursting.
This provides hints for a corona cooling process driven by soft X-rays fed by the bursts that occurred on the surface of neutron star. The flux shortage at 30--50 keV  has a time lag of 2.4$\pm$1.5 seconds with respect to that at 2--10 keV, which is comparable to that of 0.7$\pm$0.5 seconds reported in bursts of IGR 17473-2721.} We comment on the possible origin of these phenomena and on the implications for the models on the location of the corona.
\end{abstract}

\section{Introduction}
{ In low-mass X-ray binaries (LMXBs), a neutron star (\mbox{NS}) or a black hole (BH) accretes matter from a Roche-lobe-filled donor star and releases the gravitational potential-energy mostly in X-rays, either  persistently, or violently in form of the outbursts {(examples of outbursts have been shown in lower panel of Fig.1)}.} {The outbursts usually last from a few days to several months due to a sudden increase in accretion rate and vary in intensity by a factor of $ \geq  $ 100  \citep{b2}. }
During the evolution of an outburst, the system usually experiences a sequence of spectral states, which can be simplified into a high/soft (HSS) and a low/hard state (LHS). For some X-ray binaries (XRBs),
during the decay of the outburst the spectrum becomes hard again at a lower luminosity level, forming a so-called hysteresis characterized with a lagging LHS. Such hysteresis was so far observed in a
NS XRB only for IGR J17473-2721 \citep{b20}.
%
%
{ The persistent emission of the outbursts in NS LMXBS was described by the sum of a blackbody component plus a  thermal Comptonization spectrum \citep{b102}. The blackbody  component (at several keV) was attributed to the emission from the NS surface and the innermost accretion disc. The Comptonized component (at tens to hundreds of keV) was thought to originate from a hot corona. However, the location and formation process of the latter is still poorly known. According to the proposed  corona location, models for NS XRBs are divided into the Eastern model, for which the corona is in the vicinity of the NS surface \citep{b6}, and the Western model, for which the corona surrounds the disc \citep{b7}.}
Regarding the corona formation,  evaporation \citep{b5, b8, b9, b10} and magnetic re-connection models \citep{b11, b12, b13} have been put forward.


{Type-I X-ray bursts are thermonuclear explosions on the surface of neutron stars other than releasing the gravitational potential-energy.  They manifest as a rapid increase in X-ray intensity, many times brighter than the persistent emissions, followed by distinct spectral softening during bursts decay, which is the result of the cooling of the neutron star photospheres.} Unlike the timescale of days to months of the outbursts, the timescale of typical type-I bursts is tens to hundreds of seconds.  Type-I X-ray bursts, unique to NS XRBs, can be used to probe the corona. For instance, \citet{b4} observed IGR J17473-2721 during the 2008 outburst.\footnote{ { The outburst in IGR J17473-2721 had a long-lived preceding low/hard state (lasting about two months), this stable plateau represents a significant difference in comparison to the sharp peak usually  seen in atoll sources, like Aql X-1\citep{b20}.}}
They found that for bursts embedded in the preceding LHS, the 2--10 keV and 30--50 keV fluxes are anti-correlated, and that there is an apparent time lag of about 0.7 $ \pm $ 0.5  seconds present in the latter \citep{b4}.

 Here we report on LMXB 4U 1636-536 (also known as V801 Ara),  discovered with KOSO-8 \citep{b24}, and subsequently studied with SAS-3, Hakucho, Tenma, EXISAT, and RXTE \citep{b25}. { The spectral classification with the color-color diagram (CCD) has been reported by \citet{b100} and \citet{b101}}. In the X-ray CCD or hardness-intensity diagram (HID), 4U 1636-536 traces a U-shape or C-shape as a typical atoll source (detailed CCD and HID were shown in \citet{b26}). ({ The type-I bursts in Z-sources are relatively rare compared to in  atoll sources due to large accretion rate and the presence of a strong magnetic field. \citep{b1,b2} })  The orbit and spin periods of 4U 1636-536 are about 3.8 hours \citep{b16} and 581Hz \citep{b17, b18}, respectively.  Its distance was estimated as $\sim$6 kpc \citep{b19} during the type-I bursts with photospheric radius expansion (PRE).

The rest of this work is organized as follows.
Section 2  introduces the  data and the observations. Section 3 describes the findings of  additional hints for an anti-correlation between soft and hard X-rays in bursts. Under these findings, Section 4 describes the  possible constraints  upon the disk/corona configuration during state transitions. A discussion on the implications of these results is offered in Section 5.

\section{Observations and data analysis}

\begin{figure}
\centering
\includegraphics[width=3.2in]{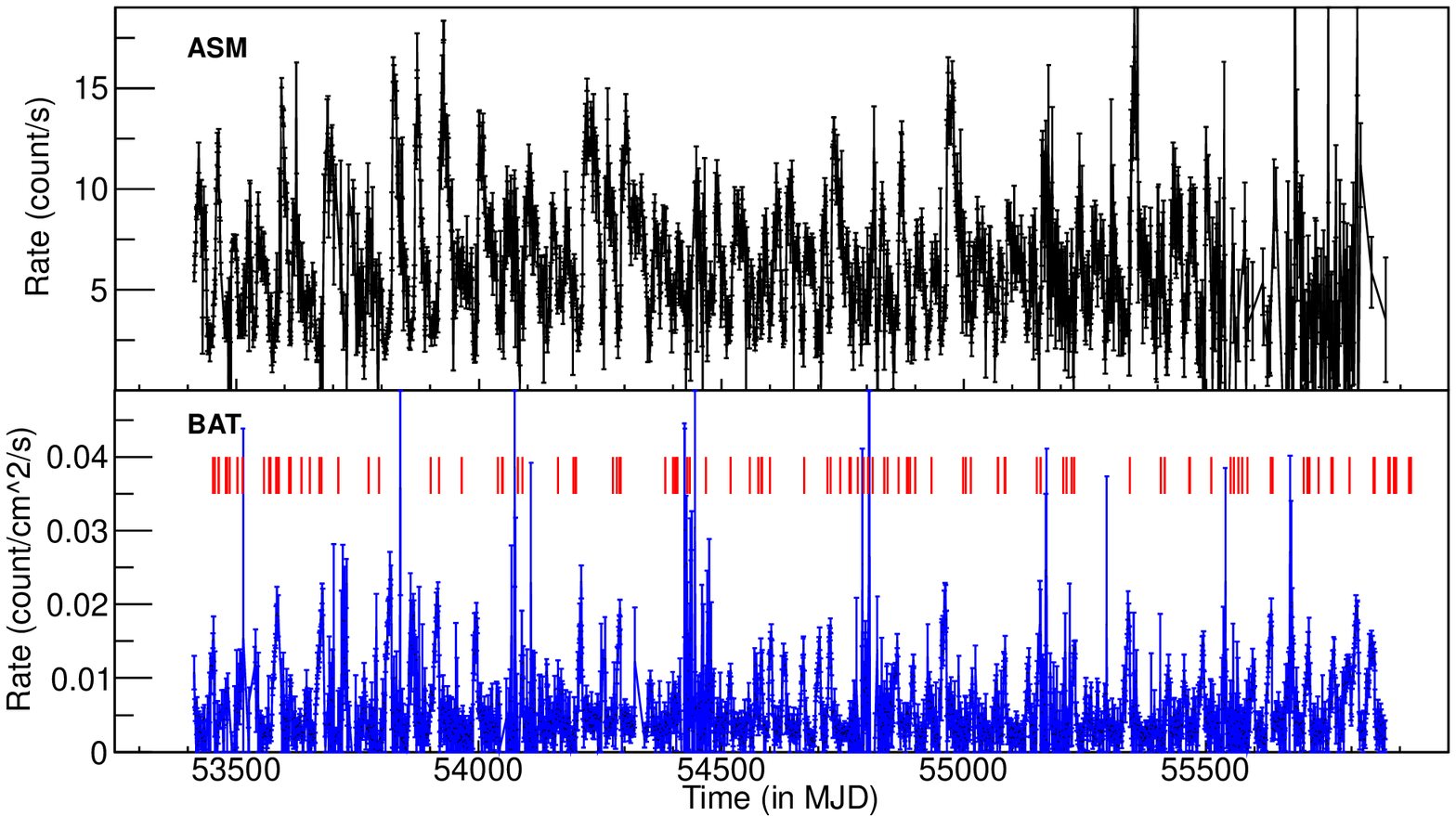}
\includegraphics[width=3.in]{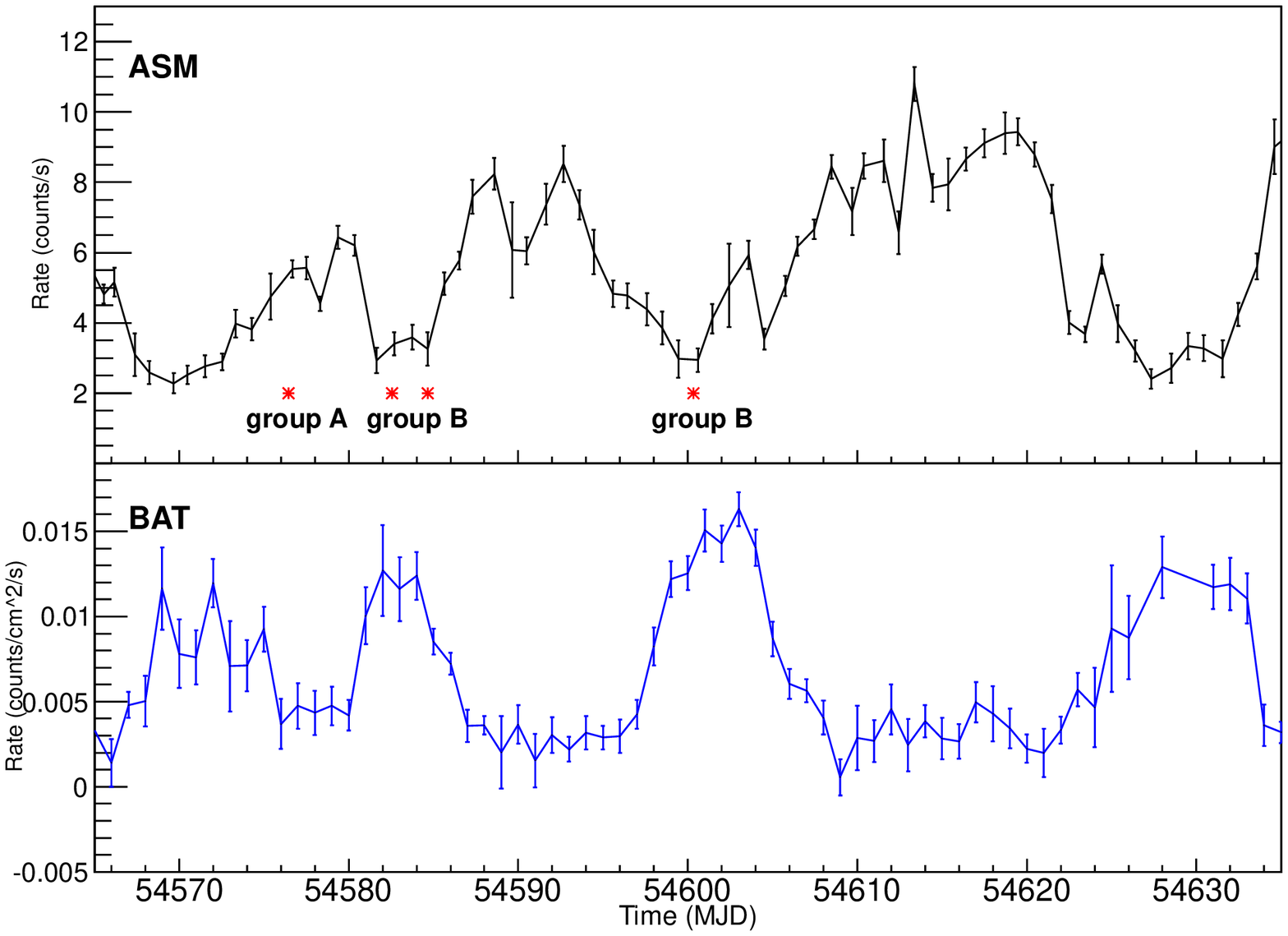}
\caption{Upper panel: the long-term monitoring of 4U 1636-536 at soft X-rays (1.5-12 keV) by ASM (black), and at hard X-rays (15-50 keV) by BAT (blue). The red lines are the type-I bursts occurred during the evolution of the source. Lower panel: example of the outbursts recorded by ASM (black) and BAT (blue). The locations of the bursts are symbolized by asterisks and classified into groups A and B.  }
\label{fig:1}
\end{figure}

4U 1636-536 has been monitored by the All Sky Monitor (ASM) of Rossi X-ray Timing Explorer (RXTE) at 1.5-12 keV since 1996. Here, we consider the RXTE observations in MJD   53452 - 55847, during which monitoring at 15-50 keV is also available from the Burst Alert Telescope (BAT) of Swift (see Fig.1).
\footnote{For more information about ASM and BAT,  see: \\http://heasarc.gsfc.nasa.gov/xte\_weather/ \\and http://swift.gsfc.nasa.gov/docs/swift/results/transients/} On top of Fig.~\ref{fig:1}, we mark 126 type-I bursts embedded in the evolution of the source. Among them, 12 PRE bursts are not included in the following analysis since their flat top peaks make it hard to co-align the individual burst according to the flux evolution. For the remaining sample of 114 burst events, we extracted their lightcurve from  the Proportional Counter Array (PCA) data \citep{b15}. We utilize the best-calibrated PCU2 only. The analysis of the PCA data is performed using HEAsoft V. 6.12. Light curves in bins of 1 second are extracted from the science event data of E\_125u\_64M\_1\_s.  The persistent emission is taken at a window of 30~s prior to each burst, and subtracted off to have the net burst lightcurve.  The dead time correction is made under the canonical criterion described at the HEASARC Web site\footnote{Please see the website: http://heasarc.nasa.gov/docs/xte/recipes\\/pca\_deadtime.html}, with the standard 1b data as a reference. Lightcurves for the bursts are co-aligned and combined according to the peak flux of each burst in the 2--10 keV band.

\section{Results}

{ The 4U 1636-536 outbursts adopted in our analysis  are shown in Fig.~1. The average peak fluxes of the outbursts are derived as 12 ct/s at 1.2--12 keV from the total ASM lightcurve,  and 0.015 ct/cm$^2$/s at 15--50 keV from the total  BAT lightcurve.}
\begin{figure}
\centering
\includegraphics[width=3.5in]{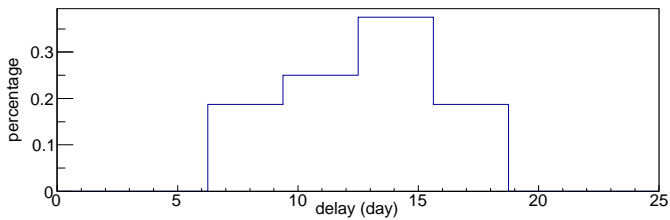}
\caption{The time delay distribution for the outbursts of 4U 1636-536. The time delay refers to the time in which the soft X-ray peak (monitored by ASM) lags the hard X-ray one (monitored by BAT).}
\label{fig:2}
\end{figure}

A  zoom-in view of a few outbursts is shown in the bottom panel of Fig.~1.
{The spectrum evolves from the low/hard state to the high/soft state during an outburst \citep{b1,b2}. Therefore, the alternately hard/soft state can be regarded as numerous separated outbursts, in which the hard X-ray peaks lead  the soft X-ray ones.}
%
%
{ We subdivided the ASM and BAT lightcurves into 30-days segments, covering in each both the individual hard  and lagging soft states, and calculated the cross correlation between them }\footnote{The cross correlation was calculated with the software CROSSCRO, and detailed information is available at http://heasarc.gsfc.nasa.gov/ftools/fhelp/crosscor.txt} and then utilized a throwing-dot method, further described below, to determine the time lag and its error. The distribution of the time lag for the outbursts is shown in Fig.~2, presenting an average of about  12.7 $\pm$ 0.2 days. {This time lag is far beyond zero which means the  hard/soft states can be distinguished clearly, and hence the type-I bursts can be well classified  accordingly.}
\begin{figure}
\centering
\includegraphics[width=3.5in]{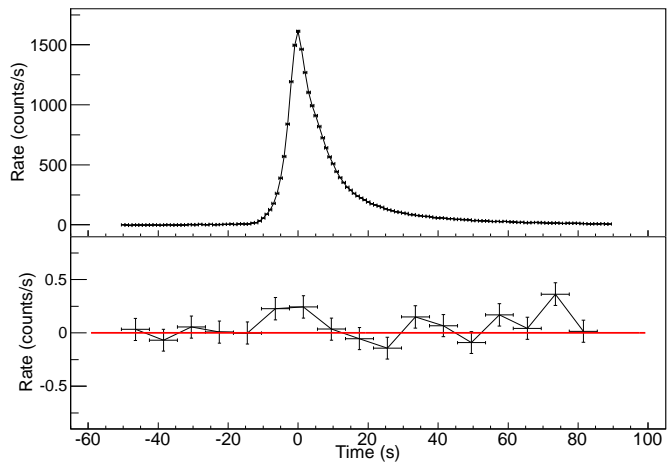}
\caption{Lightcurves derived from PCA for bursts combined in group A: 1~s-bin lightcurve  at 2--10 keV (top panel),  8~s-bin lightcurve   at 30--50 keV (bottom panel).}
\label{fig:3}
\end{figure}

\begin{figure}
\centering
\includegraphics[width=3.5in]{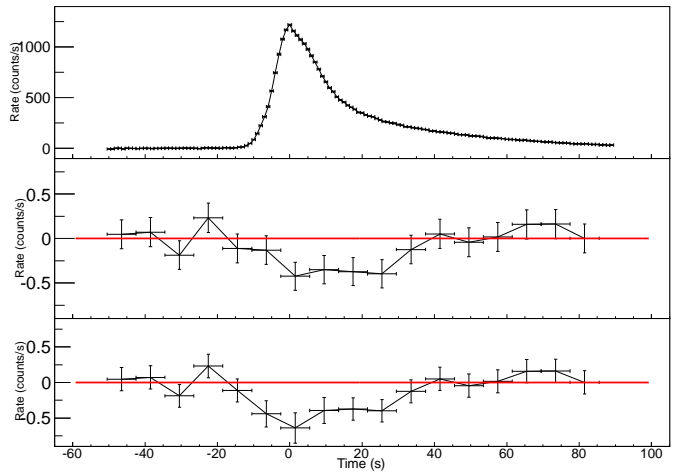}
\caption{Lightcurves derived from PCA for bursts combined in group B: 1~s-bin lightcurve  at 2--10 keV (top panel),  8~s-bin lightcurve (middle panel), and revised lightcurve (bottom panel) at 30--50 keV.}
\label{fig:4}
\end{figure}

\begin{figure}
\centering
\includegraphics[width=3.5in]{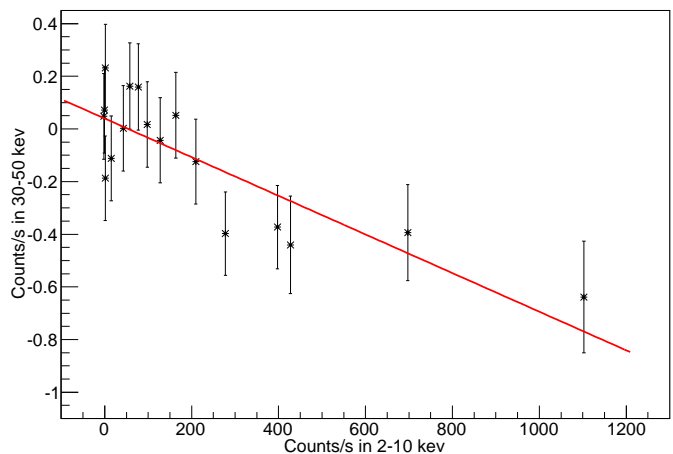}
\caption{{ Evolution for the average burst fluxes in the 2--10 keV band against the ones at the 30--50 keV band.
The 8~s-bin average burst fluxes are derived from combining the bursts enclosed in group B, which corresponds to the up and bottom panel in Fig.~4. The line shows a linear fit to the data.}}
\label{fig:5}
\end{figure}

\begin{figure}
\centering
\includegraphics[width=3.5in,angle=0]{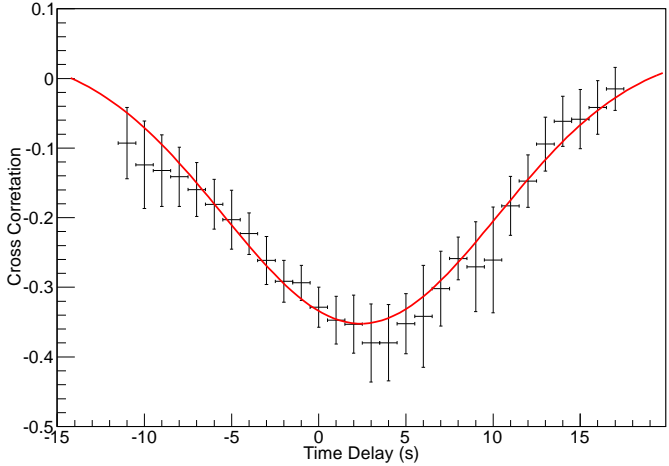}
\caption{The cross correlation between fluxes of 2--10 keV and 30--50 keV for bursts combined in group B. The fluxes are derived in bins of 1~s. The curve shows a Gaussian fit to the data for estimating the time lag in these two energy bands. The y-axis denotes the cross-correlation factor derived between the lightcurves in 2--10 and 30--50 keV bands. }
\label{fig:6}
\end{figure}

{
We subdivided the bursts into two groups.  Group A is composed of
78 bursts, which  are located around the outburst peaks in the ASM lightcurve and the each valley of the BAT lightcurve. 
Group B encloses  36 bursts embedded in the preceding LHS, with a relatively high persistent hard X-ray flux, and hence occurring in hard states of the persistent emission. The average hardness ratios (BAT flux/ASM flux) turn out to be (0.4 $\pm$ 0.1) $\times$10$^{-3}$ and (1.4 $\pm$ 0.4) $\times$10$^{-3}$ for groups A and B, respectively. Such a classification allows investigating the possibly different outburst evolution.

We combined the 2--10 keV and the 30--50 keV lightcurves for group A and averaged them in each time bin, taking the 2--10 keV peak time of each burst as reference. The resulting averaged lightcurves  are produced in bins of 1~s   at 2--10 keV and 8~s at 30--50 keV. The results are shown in Fig.~3. We notice that the 30--50 keV flux close to zero of the time axis shows a weak (statistically non-significant) excess co-aligned with the soft peak at 2--10 keV.
If it is not a statistical fluctuation,  this excess may come from the reprocessing of the burst seed photons by the corona.

We have obtained the  averaged lightcurves at 2--10 keV and 30--50 keV for group B as well
(see the top and middle panels of Fig.~4). There exists a clear shortage in the 30--50 keV flux, a  phenomenon similar to that discovered in the preceding LHS of the 2008 outburst of  IGR J17473-2721 (Chen et al. 2012). A constant fit to this lightcurve results in a $\chi^2$ of 49.40 under 25 dof, suggesting a significance of 3$\sigma$ for the shortage.\footnote{
If we assume the excess observed in Group A is real,
it can be subtracted off in the 30--50 keV flux of bursts in group B by scaling the excess derived in group A according to the difference in burst peak fluxes averaged over group A and B at 2-10 keV band. We therefore subtracted this excess off the 30--50 keV fluxes for group B and derived a revised hard X-ray lightcurve in the lowest panel of Fig.~4.}

With the 2--10 keV and the 
30--50 keV lightcurves for group B, we  investigated  the correlation along the burst evolution between the hard and soft X-rays (Fig.4).  Lightcurves in both energy bands, which corresponds to the upper and lower panel of Fig.~4, were produced with 8~s bins; the corresponding fluxes are plotted in Fig.~5. A linear fit to this flux-flux diagram gives a slope of $-(7.3\pm$1.6)$\times$10$^{-4}$ under a reduced $\chi^2$ of 0.7 (15 dof), and a correlation coefficient of $-0.8$. The  possible time delay between the two energy bands  was investigated with 1~s-bin lightcurves as well.\footnote{{ We used 1~s-bin lightcurves because for a smaller time interval, e.g. 0.5s, the statistics are insufficient to estimate time lags.}} We utilized a cross correlation and a throwing-dot method to estimate the time lag and its error, and derived a value of 2.4 $ \pm $ 1.5 seconds.
In practice, the throwing-dot method is that for bursts in group B we combined the 1~s-bin lightcurves  at 2--10 keV  and 30--50 keV, respectively, and conservatively subtracted off the excess in the 30--50 keV band, although it has little influence on the time delay. Then we sampled the lightcurve by assuming that the flux for each bin has  a Gaussian distribution, and estimated the time delay with a cross-correlation method. By sampling the lightcurve a  thousands times, the distribution of the resulted time delay was fitted with a Gaussian for inferring the error.}

\section{State transition}

{ During the outburst of an XRB, the  transition from LHS to HSS is accompanied
with a large suppression of the hard X-rays,
which is usually understood as the corona cooling by the shower of soft X-rays from the disk.
Since the type-I bursts cool the hot corona in a similar manner, and given that the luminosity and location of the type-I bursts are well known, we may use the shortage observed in hard X-rays   while bursting to constrain the corona  height  with respect to the disk at the time of the state transition. Consider the following definitions of the energy densities encounted by the corona:}
\begin{equation}
{U_{\rm burst}} = \frac{L}{{4\pi c\left( {{r_{\rm c}}^2 + {h^2}} \right)}},\\
\end{equation}
\begin{equation}
{U_{\rm disk}} = \int_{{R_{\rm in}}}^\infty  {\int_0^{2\pi } {\frac{{\sigma  \cdot T{{\left( r \right)}^4} \cdot h \cdot r}}{{\pi c{{({r^2} + r_{\rm c}^2 + {h^2} - 2{r_{\rm c}}r \cdot \cos \left( {\theta  - {\theta _{\rm c}}} \right))}^{\frac{3}{2}}}}}drd} } \theta ,\\
\end{equation}
\begin{equation}
\alpha {U_{\rm burst}} = {U_{\rm disk}},
\end{equation}
{ where $L$ is the luminosity of the type-I bursts, and $U_{\rm burst}$ and $U_{\rm disk}$ are the energy densities of the seed photons derived from the bursts and the
disk, respectively. The proportionality coefficient between them is $\alpha$. Fig.~7 sketches  the configuration of the NS-disk-corona system adopted in this estimation. The hot corona is located above the inner accretion disk, which corresponds to western  model \citep{b7}, where $r$ is the radius and $h$ is the height of the corona with respect to the disk.
%
In order to simplify the calculation, we take the multicolor blackbody model ($T{\left( {{r}} \right)^4} \propto {r}^{ - 3}$ \citep{b31}, with the innermost temperature $T_{\rm in} \sim 0.7$ keV), and the soft photons dominating the innermost disk region. We estimate the disk luminosity by approximating  it by $4\pi R_{\rm in}^2\sigma T_{\rm in}^4$. Substituting  proper values into the above equation (i.e. $L=0.7L_{\rm Edd}$, with $L_{\rm Edd}$ representing the Eddington luminosity, and an inner radius $R_{\rm in}=30$ km),  $U_{\rm burst}$ and ${U_{\rm disk}}/{\alpha}$ are shown in Fig.~8 for different values of $\alpha$ and the radius  $r$  fixed at 25 km.
With the increase of vertical height of the corona, the reduction of  $U_{\rm disk}$ is faster than that of $U_{\rm burst}$, and their intersection represents the possible height above the disk for which $U_{\rm disk}$ is sufficient for making the system transition into a HSS in an outburst.}

\begin{figure}
    \centering
    \includegraphics[width=3.5in,angle=0]{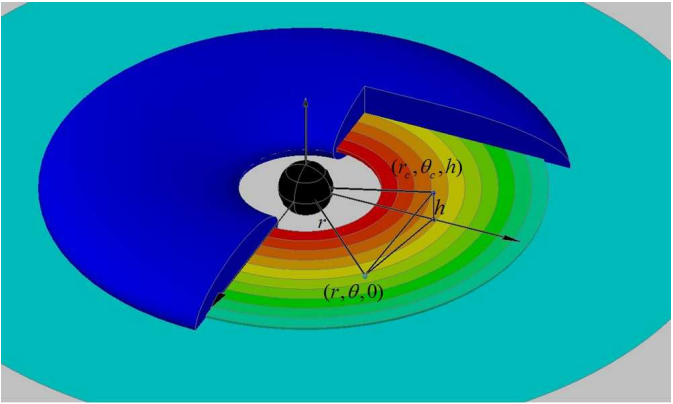}
    \caption{The NS-disc-corona system adopted in our analysis. The blue pancake represents the corona (in order to look at it more clearly, we cut a cross section). The colors in the annular rings show a cooling accretion disk at larger radii.}
    \label{fig:7}
\end{figure}

{ To further investigate the possible location of the corona while transferring into a HSS, we plot in Fig.~9 the evolution of the corona height as a function of the inner disk radius, under the condition that  $U_{\rm burst}$ and $U_{\rm disk}$ are equal. In this figure
 we take an  inversely proportional  correlation between $R_{\rm in}$ and $T_{\rm in}$, with a coefficient of 14.36 derived from the data in \cite{b23}. The correlation was derived in a form of ${T_{\rm in}} = {{14.36}}/{{{R_{\rm in}}}} + 0.22$. Here $\alpha$ is set to 1, and a constant of 0.22 is added to have $T_{\rm in} \sim 0.7$ keV at $R_{\rm in} \sim 30$ km, consistent with the parameter considerations  of Fig.~8. Fig.~9 may be regarded as a possible upper limit of the vertical height with respect to the underlying disk, beyond of which the corona would not be efficiently cooled by the disk soft X-rays. Theoretically, a corona could be held up stably  on top of the disk due to the pressure balance from gravity, magnet and gas. Here we follow   \citet{Narayan} for estimating the possible corona/disk relationship.  By assuming that a standard viscosity parameter $\alpha \sim 0.01$, and 80 percent energy releasing in the disk propagation to the hot flow, a ratio of gas pressure to magnetic pressure $\sim 2$ (weak magnetic field), then a ratio of  corona vertical height to the radial distance ($h/r$) can be approximately estimated as  0.8 \citep{Narayan}. Alternatively, a ratio of about 0.3 was derived for Sgr A in numerical calculation performed by \citet{Manmoto}. Therefore, as seen in Fig.~9, a $h/r$ value of around 0.3-0.8 would lead to a readily cooled corona  at a region near the compact object but not once further away.
We see in Fig.~9 that the disk emission is sufficient for cooling the corona once the latter is located in the vicinity of the disk. The corona is more likely cooled by disc emission at small inner radii. This is consistent with what is detected in outburst of XRBs:
changes into a HSS while the disk  moves towards the compact star. At a larger inner disk radius, the evolution shown in Fig.~9 is smoother;  in the LHS, the corona can be effectively cooled by disk emissions only if located in the vicinity of the disk. Those failed outburst as detected in the XRB H 1743-322 \citep{b99} may have kept their corona at least a few km away from the disc  during the entire outbursts to prevent an effective corona cooling.
}

\section{Discussion and Summary}


\begin{figure}
\centering
\includegraphics[width=3.5in,angle=0]{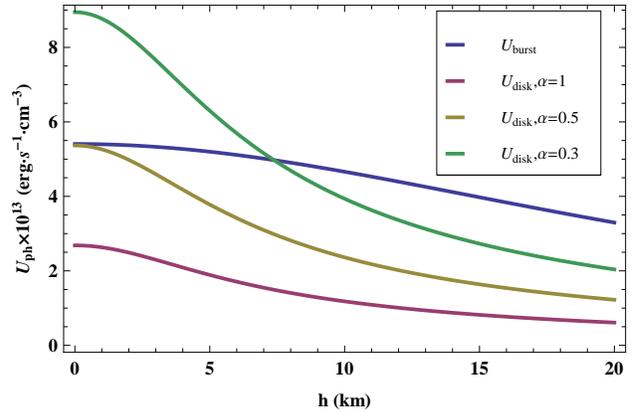}
\caption{Evolution of $U_{\rm burst}$  (${U_{\rm disk}}/{\alpha}$) as a function of the corona height with respect to the disk under different values of $\alpha$, where $U_{\rm burst}$ and $U_{\rm disk}$ are the energy densities of the seed photons derived from the bursts and the disk, respectively. The intersection shows the possible height of the corona when $r$ is fixed at 25 km.}
\label{fig:8}
\end{figure}
\begin{figure}
\centering
\includegraphics[width=3.5in,angle=0]{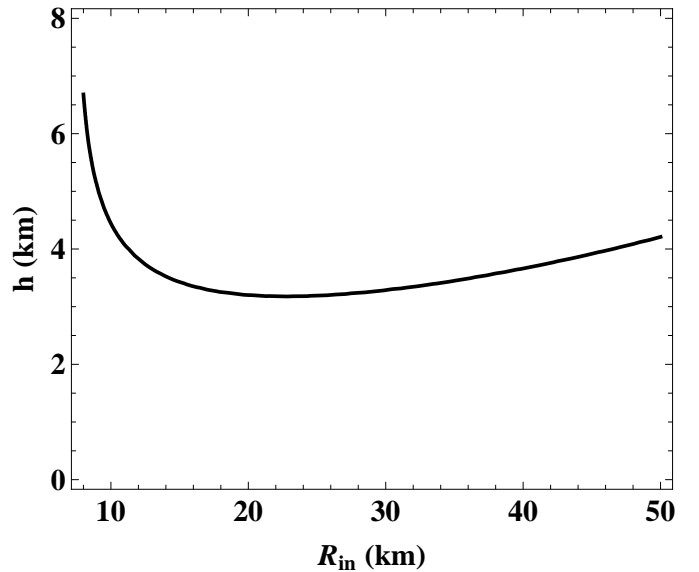}
\caption{The possible height of the corona with respect to the disk when $U_{\rm burst}$ and $U_{\rm disk}$ are equal. The vertical axis shows the corona height, and the horizontal axis represents the innermost radius of the disk during outbursts.}
\label{fig:9}
\end{figure}
{
 The shortage of hard X-ray during bursts might be a probe of evolution of an XRB outburst.
{In the LHS of an XRB, the hard X-rays  can come from a hot corona or a jet. A  jet would usually show well in the radio band.  Quantitatively,  the BH binaries are $\sim$30 times louder in radio than atoll sources at similar Eddington luminosity \citep{b27}. The correlation between $L_{\rm J}$ and $L_{\rm X}$ (the luminosity in radio and  X-rays, respectively) is quite different between NS and BH XRBs: $L_{\rm J} \propto L_{\rm X}$ in NSs, while $L_{\rm J} \propto L_{\rm X}^{0.5}$ in BHs.   This difference likely indicates that  BHs are jet-dominated while NSs are not \citep{b28,b30}.
 { As one example, \citet{b28} concluded that synchrotron self-Compton (SSC) emission alone, from the post-shock jet particles, is not a viable mechanism to explain the observed hard X-ray tail of NS 4U 0614+091. This result indicates that hot plasma exists close to the NS in the form of a corona, where the hard X-ray emission is associated with the accretion disk boundary layers.}
We thus discuss our results mainly in a context of a dominant corona-disk radiation.}

To investigate the possible cooling of the corona by soft X-ray bursts, we studied 114 bursts embedded in the X-ray
evolution of 4U 1636-536.
We subdivided these bursts in two groups, according to the hardness ratio as monitored by RXTE/ASM and Swift/BAT.
We found that in group B, when the source had harder spectrum and therefore higher flux at hard X-rays, there is a hint for an anti-correlation between the hard and the soft X-rays, similar to what was found for IGR 17473-2721.
During the bursts, the flux at 30--50 keV decreased by roughly 40\%, likely indicating that the corona was only partially cooled by the additional  soft X-rays showers fed by the bursts. As discussed in \citet{b4}, this  may suggest that the corona is not fully covering the surface of NS, but that it is perhaps located over the disk.

{The hard X-ray shortage lags that of the soft X-rays by  2.4 $ \pm $ 1.5 s. Such a time lag is comparable to the one found in IGR J17473-2721, and can be  regarded as a timescale for cooling and recovering of the corona.
It is hard to explain a corona evolution on a second timescale within a traditional disk evaporation theory \citep{b5, b8, b9, b10}. In these models, the corona is thought to be a result of the disk viscosity, and therefore its formation has a typical timescale of a few days. As discussed in \citet{b4}, an alternative model that allows such a fast evolution is magnetic field re-connection, which \citet{b11}, \citet{b12} and \citet{b13} argued for the corona formation in the solar atmosphere. The magnetic field in the inner accretion disk region can be estimated as $10^8$G \citep{b11}, consistent with that inferred  from the propeller effect \citep{b23}. However,  no characteristic magnetic field --essential for re-connection in the disk-- is indicated in current models, and the disk magnetic field remains unknown.}
}


\section*{Acknowledgments}

We acknowledge support from 973 program 2009CB824800 and the National Natural Science
Foundation of China via NSFC
11073021, 11133002, 11103020 and XTP project XDA04060604. DFT work is done in the framework of the grants AYA2009-07391, AYA2012-39303, SGR2009- 811, and iLINK2011-0303. DFT was additionally supported by a Friedrich Wilhelm Bessel Award of the Alexander von Humboldt Foundation.

\end{document}